\documentclass[fleqn,twoside]{article}
\usepackage{espcrc2}

\usepackage{graphicx}
\usepackage[figuresright]{rotating}

\newcommand{\lesssim}{\raisebox{0.3mm}{\em $\, <$} \hspace{-2.8mm}
\raisebox{-1.3mm}{\em $\sim \,$}}
\newcommand{\gtrsim}{\raisebox{0.3mm}{\em $\, >$} \hspace{-2.8mm}
\raisebox{-1.3mm}{\em $\sim \,$}}

\newcommand{\AmS}{{\protect\the\textfont2
  A\kern-.1667em\lower.5ex\hbox{M}\kern-.125emS}}

\title{Sensitivity of T2KK to non-standard interactions}

\author{Osamu Yasuda\address[MCSD]{
Department of Physics, Tokyo Metropolitan University,
Minami-Osawa, Hachioji, Tokyo 192-0397, Japan}%
}       
\begin{document}

\begin{abstract}
Assuming only the non-zero electron
and tau neutrino components $\epsilon_{ee}$,
$\epsilon_{e\tau}$, $\epsilon_{\tau\tau}$
of the non-standard matter effect and
postulating the atmospheric neutrino constraint
$\epsilon_{\tau\tau}=|\epsilon_{e\tau}|^2/(1+\epsilon_{ee})$,
the sensitivity to the non-standard interaction
in neutrino propagation of the T2KK neutrino
long-baseline experiment is estimated.
It is found that T2KK can constrain the parameters
$|\epsilon_{ee}|\lesssim 1$,
$|\epsilon_{e\tau}|\lesssim 0.2$.
\vspace{1pc}
\end{abstract}

% typeset front matter (including abstract)
\maketitle

It is expected that the undetermined
oscillation parameters such as $\theta_{13}$ and
$\delta$ are expected to be measured
in future neutrino long-baseline experiments (see, e.g.,
Ref.\,\cite{Bandyopadhyay:2007kx}).
As in the case of B factories,
such highly precise measurements will enable us to
search for deviation from the standard three-flavor oscillations.
One such possibility
is the effective non-standard neutral current-neutrino interaction
(NSI) with matter
\begin{eqnarray}
{\cal L}_{\mbox{\rm\scriptsize eff}}^{\mbox{\tiny{\rm NSI}}} 
=-2\sqrt{2}\, \epsilon_{\alpha\beta}^{fP} G_F
(\overline{\nu}_\alpha \gamma_\mu P_L \nu_\beta)\,
(\overline{f} \gamma^\mu P f'),
\label{NSIop}
\end{eqnarray}
where $f$ and $f'$ stand for fermions (the only relevant
ones are electrons, u, and d quarks),
$G_F$ is the Fermi coupling constant, and $P$ stands for
a projection operator that is either
$P_L\equiv (1-\gamma_5)/2$ or $P_R\equiv (1+\gamma_5)/2$.
In the presence of the interaction Eq.\,(\ref{NSIop}),
the standard matter effect
is modified.
Using the notation
$\epsilon_{\alpha\beta}
\equiv \sum_{P}
\left(
\epsilon_{\alpha\beta}^{eP}
+ 3 \epsilon_{\alpha\beta}^{uP}
+ 3 \epsilon_{\alpha\beta}^{dP}
\right)$,
the hermitian $3 \times3 $ matrix of the matter potential becomes
\begin{eqnarray}
{\cal A}\equiv A\left(
\begin{array}{ccc}
1+ \epsilon_{ee} & \epsilon_{e\mu} & \epsilon_{e\tau}\\
\epsilon_{\mu e} & \epsilon_{\mu\mu} & \epsilon_{\mu\tau}\\
\epsilon_{\tau e} & \epsilon_{\tau\mu} & \epsilon_{\tau\tau}
\end{array}
\right),
\label{matter-np}
\end{eqnarray}
where $A\equiv\sqrt{2}G_FN_e$.

Constraints on $\epsilon_{\alpha\beta}$
from various neutrino experiments have been discussed
by many people (see, e.g., Ref.\,\cite{Biggio:2009nt}
and references therein).
Since $\epsilon_{\alpha\beta}$ in Eq.\,(\ref{matter-np})
are given by
$\epsilon_{\alpha\beta}\sim
\sum_{P}(\epsilon^e_{\alpha\beta}
+3\epsilon^u_{\alpha\beta}
+3\epsilon^d_{\alpha\beta})$
in the case of experiments on the Earth,
we have the following constraints\,\cite{Biggio:2009nt} at 90\% CL:
\begin{eqnarray}
|\epsilon_{ee}| &<& 4\times 10^0,~~
%\nonumber\\
|\epsilon_{e\mu}| < 3\times 10^{-1},
\nonumber\\
|\epsilon_{e\tau}| &<& 3\times 10^{0\ },~~
%\nonumber\\
|\epsilon_{\mu\mu}| < 7\times 10^{-2},
\nonumber\\
|\epsilon_{\mu\tau}| &<& 3\times 10^{-1},~~
%\nonumber\\
|\epsilon_{\tau\tau}| < 2\times 10^{1\ }.
\label{current}
\end{eqnarray}

\begin{figure}[htb]
\vspace{-15mm}
\hspace{3mm}
\includegraphics[width=10.0cm]{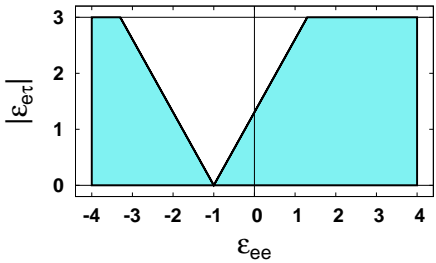}
\vspace{-15mm}
\caption{
The 90\% CL region which is
constrained by the current experimental data
in the
($\epsilon_{ee}$, $|\epsilon_{e\tau}|$) plane.
}
\label{fig1}
\end{figure}

On the other hand,
it was shown in Ref.\,\cite{Friedland:2004ah}
that
\begin{eqnarray}
|\epsilon_{e\tau}|^2
\simeq \epsilon_{\tau\tau} \left( 1 + \epsilon_{ee} \right)
\label{atm}
\end{eqnarray}
should be satisfied to be consistent with
the high-energy atmospheric neutrino data.
In the standard three-flavor scheme, the
high-energy behavior of the disappearance
oscillation probability is
\begin{eqnarray}
&{\ }&
1-P(\nu_\mu\rightarrow\nu_\mu)
%}{(\Delta E_{31}/A)^2}
\nonumber\\
&\simeq&
\left(\frac{\Delta m^2_{31}}{2AE}\right)^2\left[
\sin^22\theta_{23}
\left(\frac{c^2_{13}AL}{2}\right)^2
\right.\nonumber\\
&{\ }&\qquad\qquad+\left.
s^2_{23}\sin^22\theta_{13}\sin^2\left(
\frac{AL}{2}\right)
\right].
\label{he-std}
\end{eqnarray}
In fact it was pointed out \cite{Oki:2010uc}
that the generic matter potential
(\ref{matter-np}) leads to
the high-energy behavior of the disappearance oscillation probability
\begin{eqnarray}
1-P(\nu_\mu\rightarrow\nu_\mu)\simeq
c_0 + \frac{c_1}{E} + {\cal O}\left(\frac{1}{E^2}\right),
\label{expansion}
\end{eqnarray}
where $c_0$ and $c_1$ are functions of $\epsilon_{\alpha\beta}$,
and that $|c_0|\ll1$ and $|c_1|\ll1$ implies
$|\epsilon_{e\mu}|^2+|\epsilon_{\mu\mu}|^2+|\epsilon_{\mu\tau}|^2\ll1$
and $||\epsilon_{e\tau}|^2
-\epsilon_{\tau\tau} \left( 1 + \epsilon_{ee} \right)|\ll1$,
respectively.
Note that the terms of ${\cal O}(E^0)$
and ${\cal O}(E^{-1})$ are absent in Eq.\,(\ref{he-std})
which is in perfect agreement with the experimental data.\footnote{
So far full three flavor analysis of the atmospheric neutrino data
with the ansatz (\ref{matter-np}) has not been performed.
In Refs.\,\cite{Fornengo:2001pm,GonzalezGarcia:2004wg,Mitsuka:2008zz},
the two-flavor analysis of the atmospheric neutrino data
with the matter effect $\epsilon_{\alpha\beta}~(\alpha, \beta = \mu, \tau)$
was performed.
In Ref.\,\cite{Mitsuka:2010},
full three flavor analysis was performed, but it was
based on the assumption $\epsilon_{\alpha\beta}^{eP}=\epsilon_{\alpha\beta}^{uP}=0$.
In their analysis, therefore,
the allowed regions for the parameters
$\epsilon_{\alpha\beta}([8])\equiv 3\sum_{P}\epsilon^{dP}_{\alpha\beta}$
to be marginalized over are smaller than 
those in Eq.(\ref{current}), 
and the constraint they obtained, e.g.,
for $\epsilon_{ee}([8])\equiv 3\sum_{P}\epsilon^{dP}_{ee}$ is expected to be stronger
than that for $\epsilon_{ee}=\sum_{P}(\epsilon_{ee}^{eP}
+ 3 \epsilon_{ee}^{uP}
+ 3 \epsilon_{ee}^{dP})$.}
So, taking into account the various constraints described above,
we will work with the ansatz
\begin{eqnarray}
{\cal A}= A\left(
\begin{array}{ccc}
1+ \epsilon_{ee}~~ & 0 & \epsilon_{e\tau}\\
0 & 0 & 0\\
\epsilon_{e\tau}^\ast & 0 & ~~|\epsilon_{e\tau}|^2/(1 + \epsilon_{ee})
\end{array}
\right)
\label{ansatz}
\end{eqnarray}
in the following discussions.
The 90\% CL allowed region for the parameters
$\epsilon_{ee}$ and $|\epsilon_{e\tau}|$
is depicted in Fig. \ref{fig1}.
The region $|\tan\beta|\equiv
|\epsilon_{e\tau}/(1+\epsilon_{ee})|\gtrsim 1$
is excluded in Fig. \ref{fig1}
because of the atmospheric
neutrino data \cite{Friedland:2006pi}.

\vspace{-15mm}

\begin{figure}[htb]
\vspace{8mm}
\includegraphics[width=5.0cm]{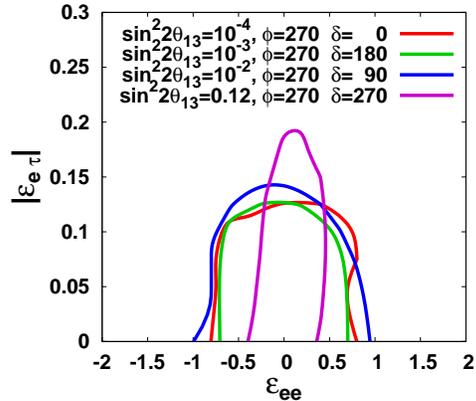}
\vspace{-10mm}
\caption{
The region which is
expected to be constrained by T2KK at 90\% CL in
the ($\epsilon_{ee}$, $|\epsilon_{e\tau}|$) plane
for various values of $\sin^22\theta_{13}$ and
for typical values of $\delta$ and
$\phi\equiv$arg($\epsilon_{e\tau}$) in degrees.
}
\label{fig2}
\end{figure}

The T2KK experiment
is a proposal for the future extension
of the T2K experiment (see, e.g., references in Ref.\,\cite{Oki:2010uc}).
In this proposal, a water Cherenkov detector
is placed both in Kamioka (at a baseline length $L$ = 295 km) and
in Korea (at $L$ = 1050 km), whereas the power of the beam
at J-PARC in Tokai Village is upgraded to 4 MW.
To examine whether T2KK can tell the existence
of NSI, we introduce
the following quantity:

\begin{eqnarray}
%&{\ }&
\Delta\chi^2  
%\nonumber \\
&=& \min_{\mbox{\scriptsize\rm param}}
 \bigg [
\sum_{i} \frac{\{ N_i({\rm NSI}) - N_i({\rm std}) \}^2}
{\sigma_i^2   }
 \nonumber \\
&{\ }&
+ \Delta\chi^2_{\mbox{\scriptsize\rm prior}}\bigg ],
\label{chi1}
\end{eqnarray}
where the prior $\Delta\chi^2_{\mbox{\scriptsize\rm prior}}$ is given by
\begin{eqnarray}
\Delta\chi^2_{\mbox{\scriptsize\rm prior}}&\equiv&
\frac{(\sin^22\theta_{23}
-\sin^22\theta_{23}^{\mbox{\scriptsize\rm best}})^2}{(\delta\sin^22\theta_{23})^2}
 \nonumber \\
 &+&\frac{(\sin^22\theta_{13}
-\sin^22\theta_{13}^{\mbox{\scriptsize\rm best}})^2}
{(\delta\sin^22\theta_{13})^2} 
 \nonumber \\
 &+&\frac{(|\Delta
 m^2_{31}|-|\Delta
 m^2_{31}|_{\mbox{\scriptsize\rm best}})^2}
{(\delta|\Delta
 m^2_{31}|)^2}.
\nonumber
\end{eqnarray}
In Eq. (\ref{chi1}) difference of the numbers of events,
$N_i({\rm NSI})$ and $N_i({\rm std})$
with or without NSI,
is compared with the error $\sigma_i$ for each bin $i$,
while $\Delta\chi^2$ is minimized with respect to
the oscillation parameters.
If $\Delta\chi^2$ is larger than, e.g., 4.6 for 2 degrees
of freedom, then significance for
the existence of NSI at
T2KK is more than 90\% CL.
In Fig.\ref{fig2}, the contour of the excluded
region in the ($\epsilon_{ee}$, $|\epsilon_{e\tau}|$) plane
at 90\% CL is plotted for various
values of $\sin^22\theta_{13}$, $\delta$ and
arg($\epsilon_{e\tau}$).
If the true point ($\epsilon_{ee}$, $|\epsilon_{e\tau}|$)
is inside each contour, then
T2KK cannot prove the existence of NSI.

\begin{figure}[htb]
\vspace{-15mm}
%\hspace{3mm}
\includegraphics[width=10.0cm]{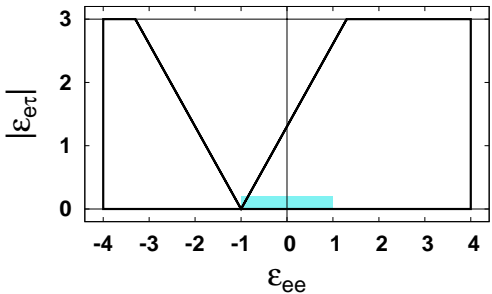}
\vspace{-15mm}
\caption{
The region which is
expected to be constrained by T2KK at 90\% CL in
the ($\epsilon_{ee}$, $|\epsilon_{e\tau}|$) plane
in the case of negative results at T2KK.
}
\label{fig3}
\end{figure}

\vglue -5mm
\noindent
We found from detailed numerical calculations \cite{Oki:2010uc}
that $|\epsilon_{ee}|\lesssim 1$,
$|\epsilon_{e\tau}|\lesssim 0.2$ will be obtained
by the negative results of the T2KK experiment.  Thus, the allowed region
in the ($\epsilon_{ee}$, $|\epsilon_{e\tau}|$) plane
will be updated from Fig.\ref{fig1} to Fig.\ref{fig3}
after the T2KK experiment is completed
with negative results.

%\vspace{-15mm}

\begin{figure}[htb]
\includegraphics[width=5.cm]{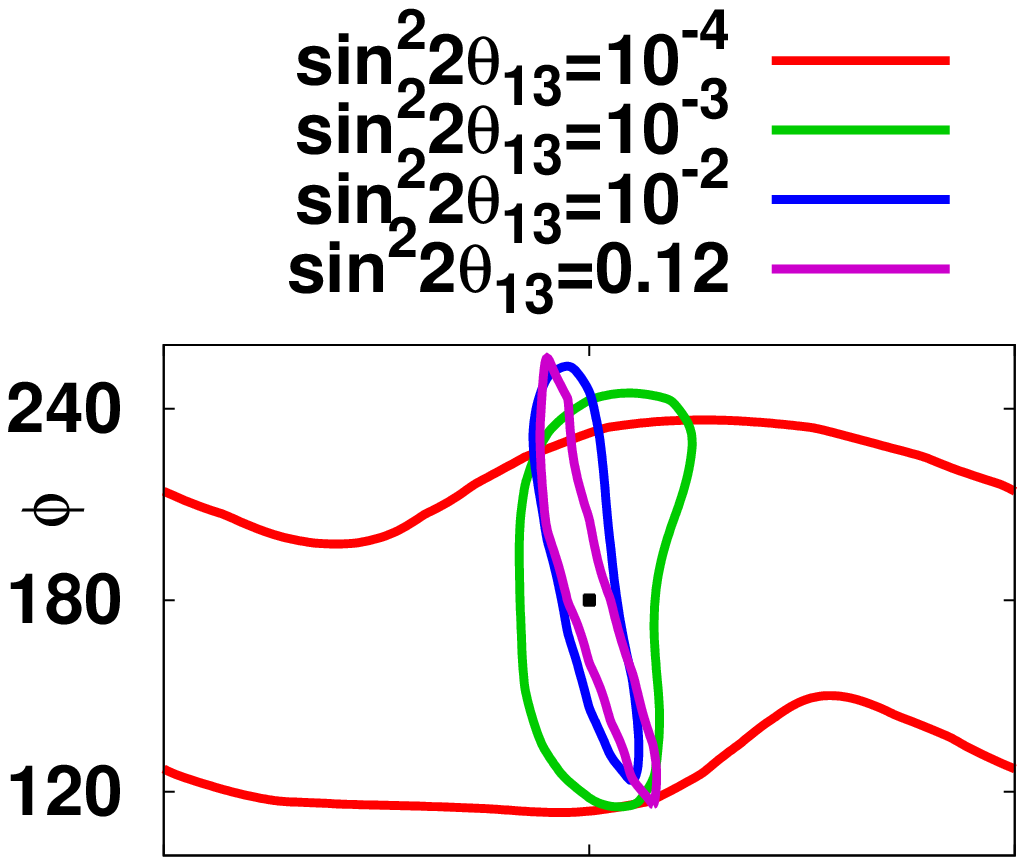}
\includegraphics[width=5.cm]{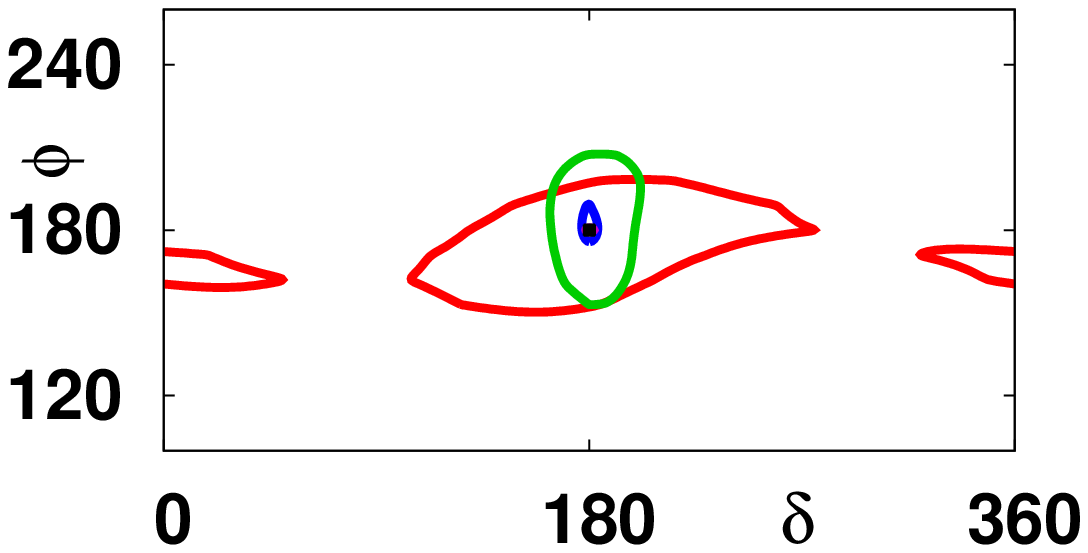}
\vspace{-10mm}
\caption{
The experimental errors at 90\% CL
in the measurement of $\delta$ and
$\phi\equiv$arg($\epsilon_{e\tau}$) at T2KK
for ($\epsilon_{ee}$, $|\epsilon_{e\tau}|$)=
(a) (0.8, 0.2), (b) (2.0, 2.0) and for various
values of $\sin^22\theta_{13}$.
The true values are
$\delta=\mbox{\rm arg}(\epsilon_{e\tau})=180$ degrees.
}
\vspace{-2mm}
\label{fig4}
\end{figure}

On the other hand, if true point
($\epsilon_{ee}$, $|\epsilon_{e\tau}|$)
is outside each contour in Fig.\ref{fig2},
then T2K should be able to obtain
some information on the parameters
$\epsilon_{ee}$, $|\epsilon_{e\tau}|$.
In this case, it becomes important whether
we can also determine the two phases
$\delta$ and arg(${\epsilon}_{e\tau}$).
The results at 90\% CL are shown in Fig.~\ref{fig4}
for (${\epsilon}_{ee}$, $|{\epsilon}_{e\tau}|$) = (0.8, 0.2)
and (2.0, 2.0).
As in the standard three-flavor case, if $\theta_{13}$ is very small,
it is difficult to get any information on $\delta$.
For larger values of $\theta_{13}$,
the sensitivity to arg($\epsilon_{e\tau}$) depends on
the value of $|\epsilon_{e\tau}|$.
For larger (smaller) values of $|\epsilon_{e\tau}|$,
sensitivity to arg($\epsilon_{e\tau}$) is good (poor).
Qualitative understanding of these features
using the analytic form of the oscillation probability $P(\nu_\mu\to\nu_e)$
is given in Ref.\,\cite{Oki:2010uc}.

In conclusion, we have studied the sensitivity of the T2KK
experiment to the non-standard
interaction in propagation.
If T2KK gets negative results, then
we have constraints $|\epsilon_{ee}|\lesssim 1$ and
$|\epsilon_{e\tau}|\lesssim 0.2$.
If T2KK obtains affirmative results, then
T2KK can determine the values of $\epsilon_{ee}$,
$|\epsilon_{e\tau}|$, and arg($\epsilon_{e\tau}$).
In particular, if the values of $\theta_{13}$
and $|\epsilon_{e\tau}|$ are relatively large
($\sin^22\theta_{13}\gtrsim {\cal O}(0.01)$,
$|\epsilon_{e\tau}|\gtrsim 0.2$), then
we can determine the two phases 
$\delta$, arg($\epsilon_{e\tau}$) separately.

I would like to thank the organizers, in particular
Prof. Fogli and Prof. Lisi, for
hospitality during the workshop.
This research was partly supported by a Grant-in-Aid for Scientific
Research of the Ministry of Education, Science and Culture, under
Grant No. 21540274.

\vspace{-7mm}

\end{document}